# SLA-Oriented Resource Provisioning for Cloud Computing: Challenges, Architecture, and Solutions

Rajkumar Buyya[1,2], Saurabh Kumar Garg[1], and Rodrigo N. Calheiros[1]

[1]Cloud Computing and Distributed Systems (CLOUDS) Laboratory
Department of Computing and Information Systems
The University of Melbourne, Australia
E-mail: {rbuyya, saurabhg, rnc}@unimelb.edu.au

[2]Manjrasoft Pty Ltd, Melbourne, Australia

*Abstract*—**Cloud computing systems promise to offer subscription-oriented, enterprise-quality computing services to users worldwide. With the increased demand for delivering services to a large number of users, they need to offer differentiated services to users and meet their quality expectations. Existing resource management systems in data centers are yet to support Service Level Agreement (SLA)-oriented resource allocation, and thus need to be enhanced to realize cloud computing and utility computing. In addition, no work has been done to collectively incorporate customer-driven service management, computational risk management, and autonomic resource management into a market-based resource management system to target the rapidly changing enterprise requirements of Cloud computing. This paper presents vision, challenges, and architectural elements of SLA-oriented resource management. The proposed architecture supports integration of market-based provisioning policies and virtualisation technologies for flexible allocation of resources to applications. The performance results obtained from our working prototype system shows the feasibility and effectiveness of SLA-based resource provisioning in Clouds.**

## I. INTRODUCTION

*"The data center is the computer,"* wrote Professor David Patterson of the University of California, Berkeley in an issue of the Communications of the ACM [1]. He noted that *"There are dramatic differences between developing software for millions to use as a service versus distributing software for millions to run their PCs."*

The above quote illustrates the challenges faced by software developers today because of recent advances in Information and Communications Technology (ICT), such as the emergence of multi-core processors and distribution of networked computing environments. These technological advances have led to the adoption of new computing paradigms which include Cloud computing, Grid computing [2], and P2P computing [3].

The next computing paradigm is envisioned to be *utility computing* [4] − the offering of computing services whenever users need them, thus transforming computing services to more commoditized utilities, similar to other utilities such as water, electricity, gas, and telephony. With this new outsourcing service model, users no longer have to invest heavily on or maintain their own computing infrastructure, and are not constrained to specific computing service providers. Instead, they just have to pay for what they use whenever they want by outsourcing jobs to dedicated computing service providers.

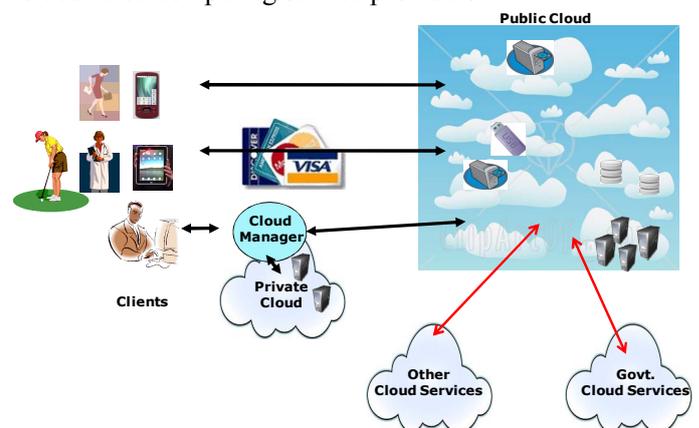

**Figure 1. Subscription-Oriented Cloud Services.**

Since users pay for using services, they want to define and expect their service needs to be delivered by computing service providers. Recently, yet another new computing paradigm called *Cloud computing* has emerged [35]. In Cloud computing, computing infrastructure and services should always be available on computing servers (which are distributed among all continents) such that companies are able to access their business services and applications anywhere in the world whenever they need to (Figure 1). Hence, Cloud computing can be classified as a new paradigm for dynamic creation of the next-generation data centers by assembling services of networked virtual machines.

To realize Cloud computing, service providers such as Amazon [5], are deploying *data centers* distributed in various locations worldwide to provide backup and ensure reliability in case of a single site failure because data centers are crucial to sustain the core business operations of companies. In addition, companies with global operations require faster response time and thus save time by distributing workload requests to multiple data centers in various locations at one time. Currently, most computing servers in data centers comprise clusters of computers as they offer high-performance, high-availability, and high-throughput processing at a lower cost compared to traditional High Performance Computing (HPC) systems. A well-known example of a company using clusters in distributed data centers is Google [6]. Google use clusters to provide replicated storage and backup servers for satisfying huge quantity of web and database requests originating from anywhere in the world.

## II. CHALLENGES AND REQUIREMENTS

Resource management issues such as SLAs (Service Level Agreements) involved in delivering software for a million users to use as a service via a data center is a lot more complex as compared to distributing software for a million users to run on their individual personal computers. There are several challenges involving SLA-*oriented resource allocation* − to differentiate and satisfy service requests based on the desired utility of users. A classification of these challenges is shown in Figure 2. By using SLAs to define service quality parameters that are required by the users from the provider, the Cloud providers know how users value their service requests and hence can provide feedback mechanisms to encourage and discourage service request submissions. In particular, finding efficient solutions for the following challenges is critical to design a fully service-oriented resource management for Cloud computing environments.

### A. Customer-driven Service Management

Yeo et al. [7] have highlighted customer satisfaction as a crucial success factor to excel in the service industry and thus proposed three user-centric objectives in the context of a computing service provider that can lead to customer satisfaction. However, there are many service quality factors that can influence customer satisfaction [20][21]. Factors that provide personalized attention to customers include enabling communication to keep customers informed and obtain feedback from them, increasing access and approachability to customers, and understanding specific needs of customers. Other factors that encourage trust and confidence in customers are security measures undertaken against risks and doubts, credibility of provider, and courtesy towards customers. Therefore, a detailed study of all possible customer characteristics needs to be done to determine if a data center needs to consider more relevant characteristics to better support SLA-oriented resource allocation.

### B. Computational Risk Management

Cloud computing is considered as the first fully accepted and implemented solution for providing computing as a utility. Having a commercial focused in offering computing services, there are several examples of elements in their resource management [7] that can be perceived as risks. For example, if SLA with a customer is violated to fulfill Quality of a request of another customer, there is a risk of penalty and customer dissatisfaction. Hence, risk analysis from the field of economics can be identified as a probable solution to evaluate these risks. However, the entire risk management process [22][23] comprises many steps and thus need to be studied thoroughly so as to fully apply its effectiveness in managing risks. The risk management process comprises the following steps: establish the context, identify the risks involved, assess each of the identified risks, identify techniques to manage each risk, and finally create, implement, and review the risk management plan.

| Architectural framework | Application requirements investigation | SLA resource allocator runtime framework | VM interaction framework | Negotiation Framework |
|---|---|---|---|---|
| SLA-based scheduling policies | Customer-driven service management | Computational risk management | Autonomic resource management | |
| SLA resource allocator | Service Request Examiner design | Admission Control design | Pricing design | Performance optimization |
| | VM Monitor design | Service Request Monitor design | Accounting design | |

**Figure 2. Challenges in SLA-based resource allocation.**

### C. Autonomic Resource Management

Service requirements of users can change over time and thus may require amendments of original service requests. As such, a data center must be able to self-manage the reservation process continuously by monitoring current service requests, amending future service requests, and adjusting schedules and prices for new and amended service requests accordingly. There are also other aspects of autonomy, such as self-configuring components to satisfy new service requirements. Hence, more autonomic and intelligent data centers are essential to effectively manage the limited supply of resources with dynamically changing service demand. For users, there can be brokering systems acting on their behalf to select the most suitable providers and negotiate with them to achieve the best service contracts. Thus, providers also require autonomic resource management to selectively choose the appropriate requests to accept and execute depending on a number of operating factors, such as the expected availability and demand of services (both current and future), and existing service obligations.

*D. SLA-oriented Resource Allocation Through Virtualization*

Recently, virtualization [24][25] has enabled the abstraction of computing resources such that a single physical machine is able to function as multiple logical VMs (Virtual Machines). A key benefit of VMs is the ability to host multiple operating system environments which are completely isolated from one another on the same physical machine. Another benefit is the capability to configure VMs to utilize different partitions of resources on the same physical machine. For example, on a physical machine, one VM can be allocated 10% of the processing power, while another VM can be allocated 20% of the processing power. Hence, VMs can be started and stopped dynamically to meet the changing demand of resources by users as opposed to limited resources on a physical machine. In particular, VMs may be assigned various resource management policies catering to different user needs and demands to better support the implementation of SLA-oriented resource allocation.

*E. Service Benchmarking and Measurement*

Recently several Cloud providers have started offering different type of computing services. Therefore competition in the IT industry is increasing to maximize their market share. From the customer perspective, it is essential to have a service measurement standard to find out most suitable services which satisfy their needs. In this context recently, Cloud Service Measurement Index Consortium (CSMIC) has identified measurement indexes (Service Measurement Index - SMI) that are important for evaluation of a Cloud service [36]. For the performance evaluation of these services, there is essential requirement of real Cloud traces from various public archives such as PlanetLab and probability distributions to model application and service requirements respectively. This is because there are currently no service benchmarks available to evaluate utility-based resource management for Cloud computing in a standard manner. Moreover, there can be different emphasis of application requirements such as data-intensive and workflow applications, and service requirements such as reliability and trust/security.

Therefore, it is necessary to derive a standard set of service benchmarks for the accurate evaluation of resource management policies. The benchmarks should be able to reflect realistic application and service requirements of users that can in turn facilitates the forecasting and prediction of future users' needs.

*F. System Modeling and Repeatable Evaluation*

The proposed resource management strategies need to be thoroughly evaluated under various operating scenarios, such as various types of resources and customers with different service requirements in order to demonstrate their effectiveness. However, it is hard and almost impossible to perform performance evaluation of resource management strategies in a repeatable and controllable manner since resources are distributed and service requests originate from different customers at any time. Hence, we will use discrete-event simulation to evaluate the performance of resource management strategies.

For our initial work [25], we have successfully used CloudSim [26] to evaluate the performance of resource management policies. CloudSim is a toolkit for modeling and simulation of Cloud resources and application scheduling. It allows easy modeling and simulation of virtual resources and network connectivity with different capabilities, configurations, and domains. It also supports primitives for application composition, information services for resource discovery, and interfaces for assigning application tasks to resources and managing their execution. Hence, these collective features in simulation toolkits such as CloudSim can be leveraged to easily construct simulation models to evaluate the performance of resource management strategies.

### III. SLA-ORIENTED CLOUD COMPUTING VISION

To meet aforementioned requirements of SLA-based resource allocation of Cloud applications, future efforts should focus on design, development, and implementation of software systems and policies based on novel SLA-oriented resource allocation models exclusively designed for data centers.

The resource provisioning within these Cloud data centers will be driven by market-oriented principles for efficient resource allocation depending on user QoS (Quality of Service) targets and workload demand patterns. In the case of a Cloud data center as a commercial offering to enable crucial business operations of companies, there are many critical QoS parameters to consider in a service request, such as reliability and trust/security. In particular, QoS requirements cannot be static and need to be dynamically updated over time due to continuing changes in business operations and operational environments. In short, there should be greater importance on customers since they pay for accessing services in data centers. The approach for realization of this research vision consists of the following:

- support for customer-driven service management based on customer profiles and QoS requirements;
- definition of computational risk management tactics to identify, assess, and manage risks involved in the execution of applications with regards to service requirements and customer needs;
- derivation of appropriate market-based resource management strategies that encompass both

- customer-driven service management and computational risk management to sustain SLA-oriented resource allocation;
- incorporation of autonomic resource management models that effectively self-manage changes in service requirements to satisfy both new service demands and existing service obligations;
- leverage of Virtual Machine (VM) technology to dynamically assign resource shares according to service requirements; and
- implementation of the developed resource management strategies and models into a real computing server in an operational data center.

IV. STATE-OF-THE-ART

Traditional Resource Management Systems (RMSs) such as Condor [8], LoadLeveler [8], Load Sharing Facility (LSF) [10], and Portable Batch System (PBS) [11], still adopt system-centric resource allocation approaches that focus on optimizing overall cluster performance. These cluster RMSs currently maximize overall job performance and system usage of the cluster. For example, they aim to increase processor throughput and utilization for the cluster, and reduce the average waiting time and response time for the jobs. They still assume that all job requests are of equal user importance and thus neglect actual levels of service required by different users. Hence, they are not able to support Cloud computing and utility computing where commercial customers' service requirements are crucial and needs to be fulfilled. Some of these batch schedulers have even extended to enable the deployment of applications across Clouds (such as EC2) dynamically. However, these resource management systems do not provide any SLA-based differentiated services to end users.

Recently, the large growth of Cloud computing led to the development of several virtual machine management platform solutions such as Eucalyptus [30] and OpenStack [12]. Apache VCL [27] is one of the first open-source systems for dynamic provision and reservation of computational resources for diverse applications within a data center through a simple web interface. Citrix Essentials [31] is a commercially available solution for infrastructure virtualization and management of data center that provides an abstraction layer between servers' hardware and the operating system. The main goal of this system is to provide automatic configuration and maintenance of data centers. Enomaly Elastic Computing Platform [29] offers a Web-based customer solution that allows users to fully control the life cycle of VMs. The Eucalyptus framework [30] is another open-source project that focuses on building IaaS clouds. The Eucalyptus provides same external interface as Amazon EC2 API and is designed to allow third-party extensions through modular software framework. The Nimbus toolkit [32] is built on top of the Globus framework having an EC2-compatible front-end API. OpenNebula [28] is also an open-source toolkit used to build private, public and hybrid clouds. Its key feature comes from scheduler module that offers dynamic resource allocation features via Haizea lease scheduler. It is possible to integrate different scheduling objectives and provisioned resources in advance. Manjrasoft Aneka [33] is a platform for building and deploying distributed applications on Clouds. It provides a rich set of APIs for transparently exploiting distributed resources and expressing the business logic of applications by using the preferred programming abstractions.

With advent of market-based resource management, many computational economy based systems [4][13] have been proposed by researchers to manage allocations of computing resources since it has been successfully adopted in the field of economics to control the supply and demand of goods. Market-based resource management is expected to regulate supply and demand of limited computing resources at market equilibrium, provide feedback in terms of economic incentives for both users and providers, and promote SLA-oriented resource allocation that differentiates service requests based on their utility and thus caters to users' needs. Therefore, market-based resource management has been proposed across numerous computing platforms that include clusters [14], distributed databases [15], Grids [16], parallel and distributed systems [17], peer-to-peer [18], and World Wide Web [19]. However, none of these market-based systems have yet considered and incorporated customer-driven service management, computational risk management, and autonomic resource management into market-based resource management, which is critical to sustain enterprise service demand in Cloud and utility computing.

In this paper, we propose a SLA-oriented resource management system built using Aneka [33] for Cloud computing. Aneka act as a market-oriented Cloud platform and allows building and scheduling of applications, provisioning and monitoring of resources with facilities such as pricing, accounting, and QoS/SLA services in private and/or public (leased) Cloud environments.

V. SYSTEM ARCHITECTURE

Figure 3 shows the high-level architecture for supporting SLA oriented resource allocation in Cloud computing [37]. There are basically four main entities involved:

- **Users/Brokers:** In general, the user interact with the Cloud management systems through an automatic systems such as brokers or schedulers who act on users behalf to submit service requests from anywhere in the world to the Clouds to be processed.

- **SLA Resource Allocator:** The SLA Resource Allocator acts as the interface between the Cloud computing infrastructure and external users/brokers. It requires the interaction of the following mechanisms to support SLA-oriented resource management:

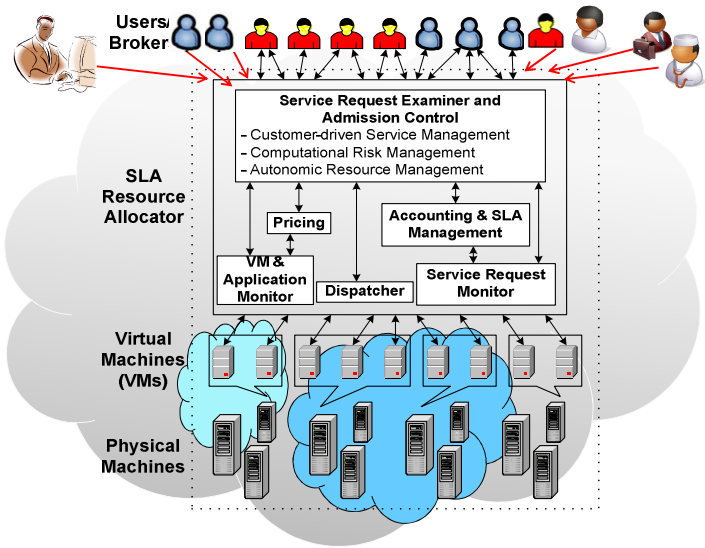

Figure 3. **High-level system architectural framework.**

  o *Service Request Examiner and Admission Control: The user* service request is first interpreted by the Service Request Examiner and Admission Control mechanism that understands the QoS requirements before determining whether to accept or reject the request. It ensures no SLA violation by reducing the chances of resource overloading whereby many service requests cannot be fulfilled successfully due to limited resources available. Therefore, it also needs the latest status information regarding resource availability (from VM Monitor mechanism) and workload processing (from Service Request Monitor mechanism) in order to make resource allocation decisions effectively. Then, it assigns requests to VMs and determines resource entitlements for allocated VMs.
    - **Autonomic Resource Management***:* This is the key mechanism that ensures that Cloud providers can serve large amount of requests without violating SLA terms. It dynamically manages the resources by using VM migration and consolidation. For instance, when an application requires low amount of resources, its VM is migrated to a host with lower capability, so that new requests can be served.

  o *Pricing*: The Pricing mechanism is a way to manage the service demand on the Cloud resources and maximize the profit of the Cloud provider. There are several ways in which service requests can be charged. For instance, requests can be charged based on submission time (peak/off-peak), pricing rates (fixed/changing) or availability of resources (supply/demand). Pricing also serves as a basis for managing computing resources within the data center and facilitates in prioritizing resource allocations effectively. Therefore, Cloud providers offer sometimes same/similar services at different pricing models and QoS levels. The two of the most prominent ones which are practically employed by Cloud providers: posted pricing and spot market.

  o *Accounting and SLA Management*: SLA Management is the component that keeps track of SLAs of customers with Cloud providers and their fulfillment history. Based on SLA terms, the Accounting mechanism maintains the actual usage of resources by requests so that the final cost can be computed and charged from the users. In addition, the maintained historical usage information can be utilized by the Service Request Examiner and Admission Control mechanism to improve resource allocation decisions.

  o *VM and Application Monitor*: Depending on the services provided, the resource management system has to keep the track of performance and status of resources at different levels. If service provided is compute resources, the VM Monitor mechanism keeps track of the availability of VMs and their resource entitlements. While in the case of application software services, the performance is continuously monitored to identify any breach in SLA and send a notification trigger to SLA Resource Allocator for taking appropriate action.

  o *Dispatcher*: The Dispatcher deploys the application on appropriate virtual resource. It also takes the responsibility of creating Virtual machine image and their initiation on selected physical hosts.

  o *Service Request Monitor:* The Service Request Monitor mechanism keeps track of the execution progress of service requests.

- **Virtual Machines (VMs):** Multiple VMs can be started and stopped dynamically to meet accepted service requests, hence providing maximum flexibility to configure various partitions of resources on the same physical machine to different specific requirements of service requests. In addition, multiple VMs can concurrently run applications based on different operating system environments on a single physical machine since every VM is completely isolated from one another on the same physical machine.

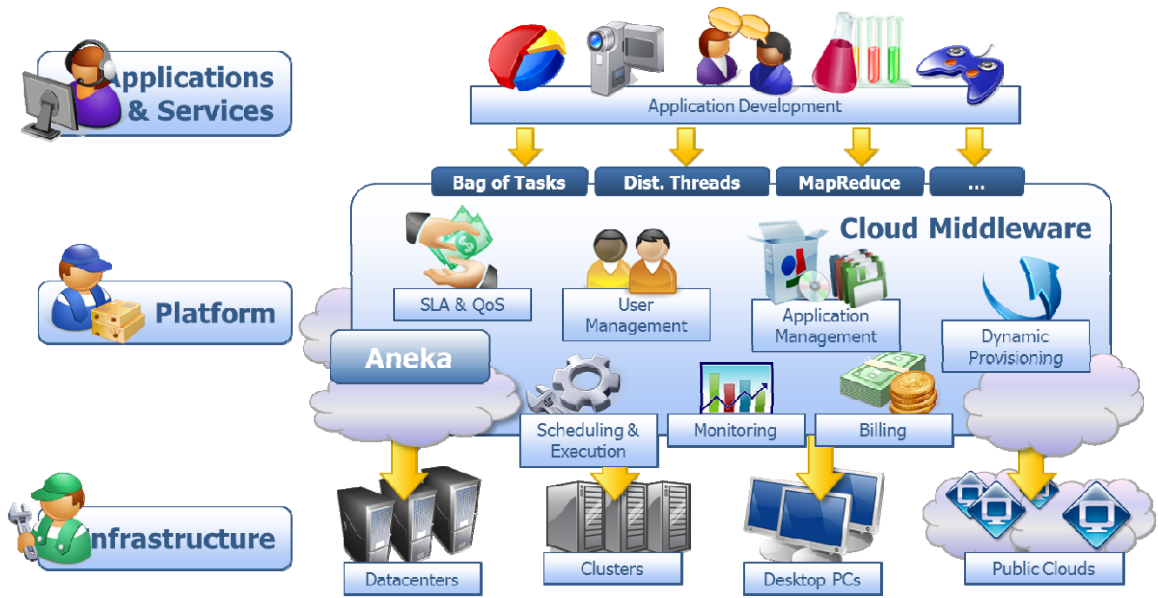

**Figure 4. Aneka architecture.**

- **Physical Machines:** The data center comprises multiple computing servers that provide resources to meet service demands.

## VI. SLA PROVISIONING IN ANEKA

To illustrate how the aforementioned architecture can be delivered in practice, we describe in this section how SLA provisioning is implemented in Aneka [33]. Figure 4 depicts the general Aneka architecture. Aneka is a Platform as a Service framework that is composed of an API for development of elastic applications supporting many programming models (Task, Thread, MapReduce), and management layer for private Clouds. The core unity of Aneka is the *Container,* which hosts one or more services such as security, scheduling, programming models, accounting, provisioning, etc. The management layer is able to handle a myriad of resources, including clusters, publics and private Clouds, and Desktop Grids.

Services in Aneka are completely decoupled from each other, and services communicate via messages sent between services. Services are designed in such a way that they are resilient to lack of reply from other services; This way, if for some reason a service is not enabled in an Aneka installation, other services keep their regular operations that are independent from the disabled services.

Figure 5 presents the Class diagram for services implementation. The interface *IService* defines all methods that an Aneka service has to implement. This interface is implemented by the abstract class *ServiceBase*, which implements methods that are common to all services, namely activation, deactivation, pausing/unpausing, and logging. *ServiceBase* also defines abstract methods that Aneka services extending *ServiceBase* have to implement. Basically, these methods are handlers for events. Moreover, different messages can also have different types of payloads, which are known to other services.

The two most important services for realization of SLA oriented framework are the Scheduler service and the Provisioning service. They both extend *ServiceBase* and define messages that are used for enabling SLA-driven execution of applications and provisioning of resources.

SLAs are defined in terms of deadline for execution of applications. The deadline, along with an estimation of execution time of each task of the application is supplied by the user during a job submission. This process is briefly described in Algorithm 1.

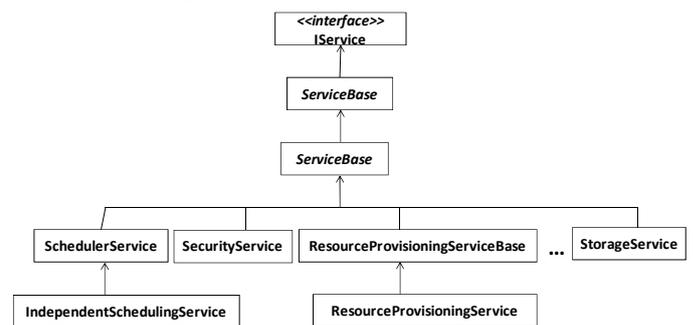

**Figure 5. Class diagram of Aneka services.**

Aneka handles the process of dynamic provisioning to meet user SLAs in autonomic manner. Additional resources are provisioned for applications when required and are removed when they are not necessary. When the Scheduler service receives a completed task or a new job, it estimates if the deadline can be achieved with the current number of resources allocated to the job. Moreover, because the initial estimation given by the user about execution time of tasks may be inaccurate, estimation is updated with data obtained from the actual execution of tasks.

> **Algortihm 1. SLA-oriented Dynamic Provisioning Algortihm in Aneka.**
>
> 1. When a task finishes or a new job is received:
>    1.1. Updates estimation of task runtime;
>    1.2. Defines estimated job completion time with current amount of resources;
>    1.3. If completion time > deadline
>        1.3.1. Determines number of extra resources required
>        1.3.2. Submits a request for resources to the Provisioner.
>    Else
>        1.3.2. If resources can be released
>            1.3.2.1. Submits request for release of resources to the Provisioner

If additional resources are not required, the Scheduler checks if some dynamically allocated resources can be removed from the job. If the Scheduler decides whether the deadline can be met with fewer resources considering the updated estimation of task runtime; the excessive resources are removed and made available to other jobs (or decommissioned, as described below). A detailed description of the Resource provisioning process in Steps 1.3.1 and 1.3.2 from the Algorithm 1 is available in our previous work [34]. The required number of resources is passed to the *Dynamic Provisioning* service, which acquires these resources from external sources such as public Clouds.

Figures 6 and 7 illustrate the iteration of diverse Aneka services in order to meet application's QoS and the diverse states a job can experience during their lifetime. When a QoS-enabled job is received by Aneka, it goes to the QoS state, which means that tasks that compose the job will have higher scheduling priority than non QoS-enabled tasks. Both QoS and regular tasks can go to Queued state if no resources are promptly available to them. However, for the case of QoS-enabled tasks extra resources will be provisioned. When the Scheduler service detects that extra resources are required for a job, either because the QoS-enabled job is queued or because the estimation is such that deadline cannot be met, the job goes to the underprovisioned state. If deadline is achievable with available resources, the job goes to the provisioned state.

Finally, if deadline cannot be met (for example, because deadline is smaller than the boot time of provisioned resources), the job goes to unfeasible state.

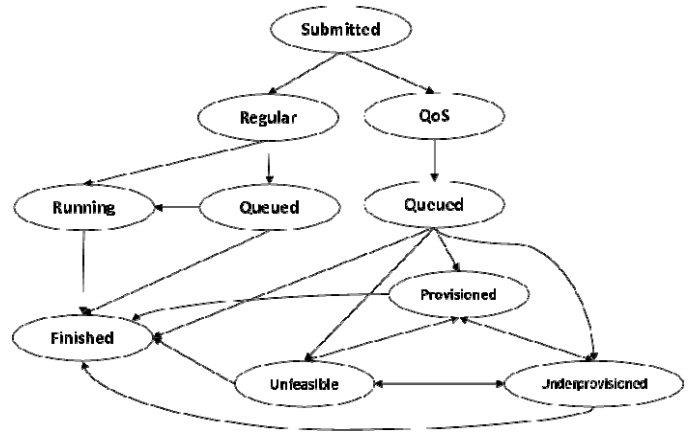

**Figure 6. State diagram of jobs in Aneka.**

When the job goes to the underprovisioned state, a message is sent from the Scheduler service to the Dynamic Provisioner service. This message contains the number of resources required by the job. The scheduler is not aware of possible sources from resources for the tasks; decision on the suitable resource pool is done by the Resource Pool Manager. The Resource Pool Manager returns to the Scheduler references for the new resources. However, these resources will only be able to receive tasks for execution when the corresponding Aneka worker starts running and announces its availability to the Scheduler.

When extra resources become available, feasibility of the corresponding underprovisioned job is recalculated. From that, it can go to the state *provisioned*, if enough resources are available, it can stay in the *underprovisioned* state, in which case it can receive more dynamically provisioned resources, or it can go to the unfeasible state. Jobs in the latter stage continue to receive dynamic provisioned resources, but no extra allocations are made for them. If for some reason an unfeasible job returns to *underprovisioned* state (for example, because some tasks completed before the expected time), it may be candidate to receive extra dynamically provisioned resources. Also, changes in state may happen every time a task completes execution and the estimation of finish time for the job is recalculated.

For both regular and QoS-enabled jobs, *finished* state is reached in three different situations: (i) execution of all the tasks finished; (ii) execution of the job failed; and (iii) the user cancelled the job.

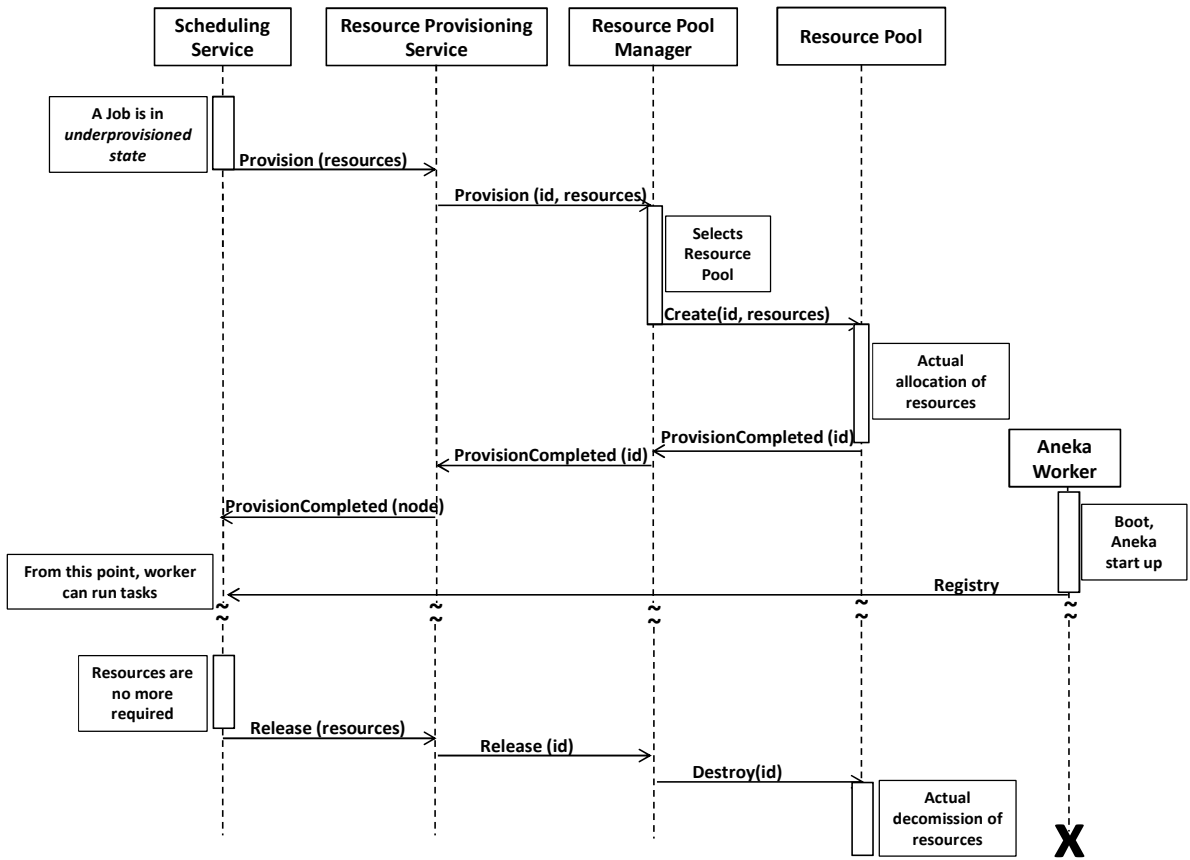

Figure 7. Interaction between Aneka services during dynamic provisioning process.

In the case of resources acquired from public Cloud providers, there is typically a time window for which the utilization of the resource is paid. In this case, resources are not decommissioned before the end of such time window, in order to maximize utilization of such resources. Therefore, when a QoS-enabled job achieves the *finished* state, the corresponding resources are evaluated for the available time before their next billing period. Those resources that can be used are made available for other QoS-enabled jobs. If there is no other QoS-enabled jobs, resources are made available for regular tasks until the end of the billing period. When the end of billing period of a resource is achieved, a check is made to verify if removing such resource will cause a QoS-enable task to go to *underprovisioned* state. If this situation is detected, resource is not decommissioned.

Notice that the Scheduler is not the Aneka service that is responsible for decommissioning resources; when it detects the resource can be decommissioned, a message is sent to the Resource Pool to proceed with the decommission process. This is because procedure for decommission varies depending on the specific source of resources, and these different sources are associated to different Resource Pools. The Resource Pool Manager is the service that keeps track of available resource pools and that is able to interact with them for executing resource management tasks. This mechanism enables execution of jobs inside a user-defined deadline. An evaluation of such a mechanism is presented in the next section.

VII. PERFORMANCE EVALUATION

The evaluation of the SLA provisioning mechanism of Aneka, described in the previous section, has been carried out entirely in Amazon EC2, USA East Coast.

The experimental setup consists of static resources and dynamic resources. Static resources are composed of 5 machines. One machine, running the Aneka master, is an *m1.large* (7.5 GB of memory, 4 EC2 Compute Units, 850 GB of local instance storage, 64-bit platform, U$0.48 per instance per hour) Windows-based virtual machine. The other 4 machines, which are Aneka workers, are *m1.small* (1.7 GB of memory, 1 EC2 Compute Unit, 160 GB of local instance storage, 32-bit platform, U$0.085 per instance per hour) Linux-based virtual machines. Dynamic resources provisioned are of type *m1.small* Linux-based instances.

A CPU-intensive application is used for experiments. SLA is defined in terms of user-defined deadline. For the purpose of this experiment, execution time of each task was set to 2 minutes. Each job consists of 120 tasks. Therefore, the total execution time of the job in a single machine is 4 hours.

We executed such a job initially without any QoS configuration. Afterwards, we repeated the experiment with different deadlines set for the job: 45 minutes, 30 minutes, and 15 minutes. The results for execution of the job without QoS and with different deadlines are given in Table 1.

**Table 1. Experimental results.**

|        | Static machines | Dynamic machines | Execution Time | Extra cost |
|--------|-----------------|------------------|----------------|------------|
| No QoS | 4               | 0                | 1:00:58        | 0          |
| 45min  | 4               | 2                | 0:41:06        | U$ 0.17    |
| 30 min | 4               | 6                | 0:28:24        | U$ 0.51    |
| 15 min | 4               | 20               | 0:14:18        | U$ 1.70    |

The results show that Aneka can effectively meet QoS-requirements of applications by dynamically allocating resources. They also show that the provisioning algorithm of Aneka performs cost-optimization: it allocates the minimum amount of resources that enable the deadline to be met. This is evidenced by the fact that execution times were very close to the deadline. Another possible strategy, time-based optimization, would adopt a more aggressive dynamic provisioning utilization in order to reduce execution time. This would allow deadlines to be met by larger margins than the obtained with the current strategy, but would incur in more cost for the users. Effective policies for time-based optimization will be subject of future research and development of Aneka.

## VIII. CONCLUSIONS AND FUTURE DIRECTIONS

In the next twenty years, service-oriented computing will play an important role in sharing the industry and the way business is conducted and services are delivered and managed. This paradigm is expected to have major impact on service economy; the service sector includes health services (e-health), financial services, government services, etc. This involves significant interaction between clients and service providers. With increased dependencies on ICT technologies in their realization, major advances are required in user QoS-based allocation of resources to competing applications in a shared environment provisioning though a secure virtual machines.

In this paper, we pointed out many challenges in addressing the problem of enabling SLA-oriented resource allocation in data centers to satisfy competing applications demand for computing services. In particular, the user applications are becoming more complex and need multiple services to execute instead of a single service. These complex user applications often require collaboration among multiple organizations or businesses and thus require their specific services to function successfully. Moreover, fast turnaround time is needed for running user applications in today's increasingly competitive business environment. Hence, by addressing SLA-oriented resource allocation in data centers, we provide a critical link to the success of the next ICT era of Cloud computing. We also showed how the proposed framework can be effectively implemented using the Aneka platform.

We envision the need for a deeper investigation in SLA-oriented resource allocation strategies that encompass customer-driven service management, computational risk management, and autonomic management of Clouds in order to improve the system efficiency, minimize violation of SLAs, and improve profitability of service providers.

ACKNOWLEDGEMENTS

We thank our colleagues who contributed towards the development of Aneka platform, especially Christian Vecchiola, Dileban Karunamoorthy, and Xingchen Chu. We thank Chee Shin Yeo for his earlier contribution to the work reported in this paper. We thank Deepak Poola and William Voorsluys for their comments on the paper.